\documentclass[journal=jacsat,manuscript=article]{achemso}

\usepackage[utf8]{inputenc}
\usepackage[T1]{fontenc}
\usepackage{mathptmx}
\usepackage{xcolor}
\usepackage[version=3]{mhchem} % Formula subscripts using \ce{}

\author{Bruno H. S. Mendon\c{c}a}
\affiliation[UFMG]{Departamento de F{\'i}sica, ICEX, Universidade Federal de Minas Gerais, CP 702, Belo Horizonte 30123-970, MG, Brazil}
\email{brunnohennrique13@gmail.com}
\author{Elizane E. de Moraes}
\affiliation[UFBA]{Instituto de F{\'i}sica, Universidade Federal da Bahia, Campus Universit{\'a}rio de Ondina, Salvador 40210-340, BA, Brazil}
\altaffiliation{Catalan Institute of Nanoscience and Nanotechnology (ICN2), CSIC and BIST, Campus UAB, Bellaterra, 08193 Barcelona, Spain}

\author{João P. K. Abal}
\affiliation[UFRGS]{Instituto de F{\'i}sica, Universidade Federal do Rio Grande do Sul, Porto Alegre 91501-970, RS, Brazil}

\author{João V. L. Valle}
\affiliation[UFBA]{Instituto de F{\'i}sica, Universidade Federal da Bahia, Campus Universit{\'a}rio de Ondina, Salvador 40210-340, BA, Brazil}

\author{T{\'a}ssylla O. Fonseca}
\affiliation[UFMG]{Departamento de F{\'i}sica, ICEX, Universidade Federal de Minas Gerais, CP 702, Belo Horizonte 30123-970, MG, Brazil}

\author{H{\'e}lio Chacham}
\affiliation[UFMG]{Departamento de F{\'i}sica, ICEX, Universidade Federal de Minas Gerais, CP 702, Belo Horizonte 30123-970, MG, Brazil}

\title[An \textsf{achemso} demo]
  {Influence of carbon nanocone structure on ultra-efficient water flow}

\abbreviations{IR,NMR,UV}

\begin{document}

%%%%%%%%%%%%%%%%%%%%%%%%%%%%%%%%%%%%%%%%%%%%%%%%%%%%%%%%%%%%%%%%%%%%%
%% The "tocentry" environment can be used to create an entry for the
%% graphical table of contents. It is given here as some journals
%% require that it is printed as part of the abstract page. It will
%% be automatically moved as appropriate.
%%%%%%%%%%%%%%%%%%%%%%%%%%%%%%%%%%%%%%%%%%%%%%%%%%%%%%%%%%%%%%%%%%%%%
%\begin{tocentry}

%Some journals require a graphical entry for the Table of Contents.
%This should be laid out ``print ready'' so that the sizing of the
%text is correct.

%Inside the \texttt{tocentry} environment, the font used is Helvetica
%8\,pt, as required by \emph{Journal of the American Chemical
%Society}.

%The surrounding frame is 9\,cm by 3.5\,cm, which is the maximum
%permitted for  \emph{Journal of the American Chemical Society}
%graphical table of content entries. The box will not resize if the
%content is too big: instead it will overflow the edge of the box.

%This box and the associated title will always be printed on a
%separate page at the end of the document.

%\end{tocentry}

%%%%%%%%%%%%%%%%%%%%%%%%%%%%%%%%%%%%%%%%%%%%%%%%%%%%%%%%%%%%%%%%%%%%%
%% The abstract environment will automatically gobble the contents
%% if an abstract is not used by the target journal.
%%%%%%%%%%%%%%%%%%%%%%%%%%%%%%%%%%%%%%%%%%%%%%%%%%%%%%%%%%%%%%%%%%%%%
\begin{abstract}
In this study, using nonequilibrium molecular dynamics simulation, the water flow in carbon nanocones is studied using the TIP4P/2005 rigid water model. The results demonstrate a nonuniform dependence of the flow on the cone apex angle and the diameter of the opening where the flow is established, leading to a significant increase in the flow in some cases. The effects of cone diameter and pressure gradient are investigated to explain flow behavior with different system structures. We observed that some cones can optimize the water flow precisely. Nanocones with a larger opening facilitate the sliding of water, significantly increasing the flow, thus being promising membranes for technological use in water impurity separation processes. Nanocones with narrower opening angles limited water mobility due to excessive confinement. This phenomenon is linked to the ability of water to form a larger hydrogen-bond network in typical systems with diameters of this size, obtaining a single-layer water structure. Nanocones act as selective nanofilters capable of allowing water molecules to pass through while blocking salts and impurities. The conical shape of their structures creates a directed flow that improves separation efficiency. Membranes based on carbon nanocones are becoming promising for clean, smart, and efficient technologies. The combination of transport speed, selectivity, and structural control put them ahead of other nanostructures for various purposes.
\end{abstract}

%%%%%%%%%%%%%%%%%%%%%%%%%%%%%%%%%%%%%%%%%%%%%%%%%%%%%%%%%%%%%%%%%%%%%
%% Start the main part of the manuscript here.
%%%%%%%%%%%%%%%%%%%%%%%%%%%%%%%%%%%%%%%%%%%%%%%%%%%%%%%%%%%%%%%%%%%%%
\section{Introduction}

The behavior of water at the nanoscale has recently attracted significant interest for engineering applications such as water desalination and purification. This interest stems from the enhanced interfacial effects of nanofluidic devices, where most fluid molecules interact with the surface \cite{10.1016@j.carbon.2017.12.039,@10.1021/acs.jpcb.3c02889,@10.1063/1.5086054,mendoncca2023water,valle2024accuracy,10.1063/5.0243115,yasmeen2025enhanced,10.1039/D4CP01068J,zhang2024exploring,tan2024novo,10.1016/j.physa.2018.11.042,10.1016/j.chemphys.2020.110849,10.1063/5.0031084,10.1063/1.5129394,10.1021/acs.jpcc.4c00078}. Consequently, constructing these devices from materials that facilitate rapid and efficient water transport is paramount. In the past decade, advances in the manipulation and manufacturing of nanomaterials have greatly improved, aiding in the identification of suitable materials and channel geometries and sizes that enhance water transport \cite{@10.1021/nl200843g,10.1080/1536383X.2024.2445668,abal2021molecular,yasmeen2025enhanced,10.1016/j.cej.2024.158366,10.1017/flo.2024.34,10.1039/D4CP01068J,10.1126/sciadv.adj3760,10.1002/smll.202409950}. Specifically, carbon allotropes, such as carbon nanotubes, graphite, and graphene, have emerged as potentially important materials in this field \cite{qiao2003ion,10.1016/j.inoche.2018.02.011,singh2021nanomembranes,10.1016/j.inoche.2018.02.011,10.1038/s41598-018-26072-6,10.1002/9783527848546.ch4,10.1016/j.molliq.2024.126705,li2024electric,li2024molecular}.

The phenomenon of enhanced water transport—often referred to as superflow—was notably reported by Qin et al.\cite{@10.1021/nl200843g}, who demonstrated that water confined in carbon nanotubes can flow up to 900 times faster than predicted by classical hydrodynamics. This discovery has spurred several research studies into low-friction nanomaterials such as graphene, graphitic surfaces, boron nitride nanotubes, and other nanomaterials \cite{suk2010water,li2023modulation,abal2021molecular,@10.1021/acs.jpcb.3c02889,@10.1002/aic.17543,li2024molecular,@10.1002/aic.17543,10.1016@j.carbon.2017.12.039,10.1103/PhysRevLett.102.184502,kargar2019water,zhou2024interlink,li2024electric,yang2023fast,10.1103/PhysRevE.111.L023101}. Among these, carbon nanocones (CNCs) have attracted increasing attention due to their sp$^{2}$ hybridization of graphene with a distinct conical morphology \cite{10.1088/1468-6996/10/6/065002,charlier2001electronic,10.1038@41284, @10.1016/j.jmgm.2025.109018,mistry2021untangling,chen2022slip,feng2016single,@10.1016/j.comptc.2022.113976}, which naturally resembles the hourglass-shaped structure of aquaporins (AQPs) - the biological channels responsible for highly efficient and selective water transport across cell membranes \cite{@10.1002/anie.200460804, 10.1063/1.4893782}. In AQPs, the narrow neck ensures selectivity, while the conical entrance reduces hydrodynamic resistance and improves permeability \cite{10.1063/1.4893782,@10.1073/pnas.0902725106,@10.1002/anie.200460804,@10.1063/1.4897253}. Inspired by this natural design, CNCs offer a biomimetic pathway to achieve high water flux and selectivity in synthetic systems \cite{@10.1002/aic.17543}. Recently, molecular dynamics studies by Leivas et al. \cite{leivas2023atmospheric,@10.1063/5.0142718} demonstrated that hydrophilic carbon nanocones can efficiently capture water from both liquid and vapor phases, with enhanced performance under vapor exposure and scalability dependent on intercone spacing. In this context, the asymmetric shape of CNTS can also induce directional flow and reduce energy barriers, making them attractive candidates for nanofluidic devices and high-performance desalination membranes.

In this work, we performed a systematic study to investigate the effect of the structure of carbon nanocones on the water flow properties, using the classical approach of nonequilibrium molecular dynamics simulations. The water flow was related to the cones' opening angle, hydrogen bond breaking/forming capacity, flow, and the influence of the external pressure gradient to which the filtration system is subjected. The remainder of this manuscript is organized as follows: In Sec. II, we present the simulation methodology used in this analysis and define the simulated models. In Sec. III, we discuss the results, and in Sec. IV, we present the conclusions.

%%%%%%%%%%%%%%%%%%%%%%%%%%%%%%%%%%%%%%%%%%%%%%%%%%%%%%%%%%%%%%%%%%%%%
\section{Simulation methodology}

\begin{figure}[H]
	\begin{center}
		\includegraphics[width=5.4in]{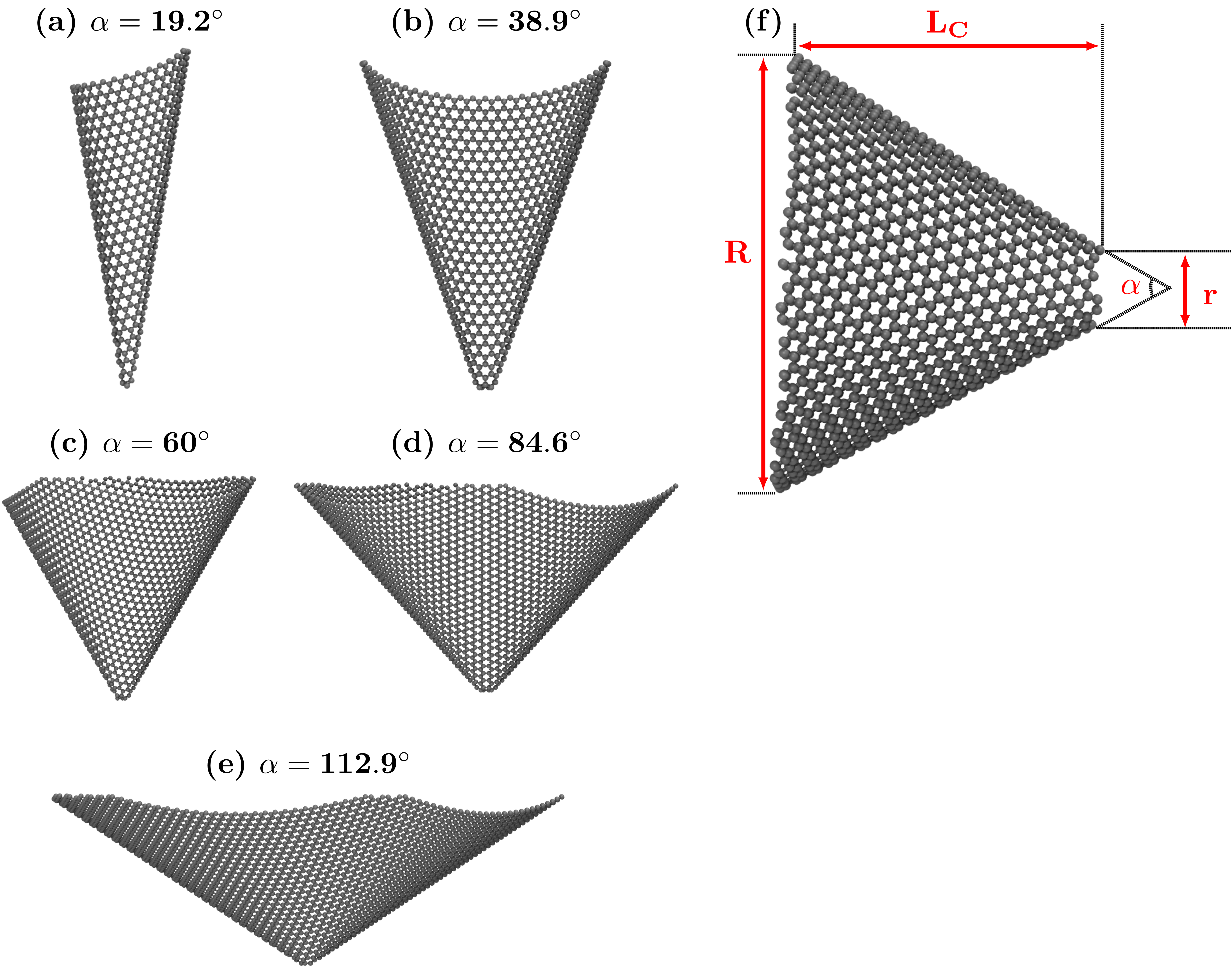}
	\end{center}
	\caption{Snapshot of the carbon nanocones used in the simulations, with the following opening angles (a) 19.2$^{\circ}$, (b) 38.9$^{\circ}$, (c) 60$^{\circ}$, (d) 84.6$^{\circ}$, (e) 112.9$^{\circ}$ e (f) schematic showing of the parameters of the nanocone dimensions of interest.} 
	\label{fig_cncs}
\end{figure}

We used the non-equilibrium molecular dynamics method~\cite{10.1063@1.1743957,10.1063@1.1730376} to simulate the transport of water from one reservoir through a carbon nanocone (CNC) membrane to another reservoir~\cite{10.1063@1.5086054,10.1063@5.0039963,10.1021@acsanm.1c01982,10.1039@D1CP00613D,10.1039@D0CP00484G}. The developed water reservoir system was based on the model proposed by Huang et al.~\cite{10.1063@1.2209236}. CNCs are conical structures that can theoretically be built by removing a sector from a circular graphene sheet, and reattaching the sheet in a conical shape. We considered, for the construction of the membranes, the five CNCs observed experimentally~\cite{10.1038@41284}, with the following opening angles 19.2$^{\circ}$, 38.9$^{\circ}$, 60$^{ \circ}$, 84.6$^{\circ}$ and 112.9$^{\circ}$, shown in Figure~\ref{fig_cncs}. The length of each nanocone $L_{C}$, as well as the number of pentagons $N_{P}$, the diameters $R$ and $r$ and the opening angle $\alpha$, are detailed in Table~\ref{tbl:cncs}. 

\begin{table}[H]
	\caption{Parameters for the used carbon nanocones.}
	\label{tbl:cncs}
	\begin{tabular}{lllll}
		\hline
		$N_{P}$ & $\alpha$ & $L_{C}$ (nm) & $R$ (nm) & $r$ (nm)  \\
		\hline
		1       & 112.9$^{\circ}$ & 4.37 & 8.08 & 0.86 \\
		2       & 84.6$^{\circ}$  & 4.31 & 6.55 & 0.82 \\
		3       & 60$^{\circ}$    & 4.28 & 5.09 & 0.81 \\
		4       & 38.9$^{\circ}$  & 4.26 & 3.59 & 0.76 \\
		5       & 19.2$^{\circ}$  & 4.43 & 2.17 & 0.76 \\
		\hline
	\end{tabular}
\end{table}

\begin{figure}[H]
	\begin{center}
		\includegraphics[width=6.4in]{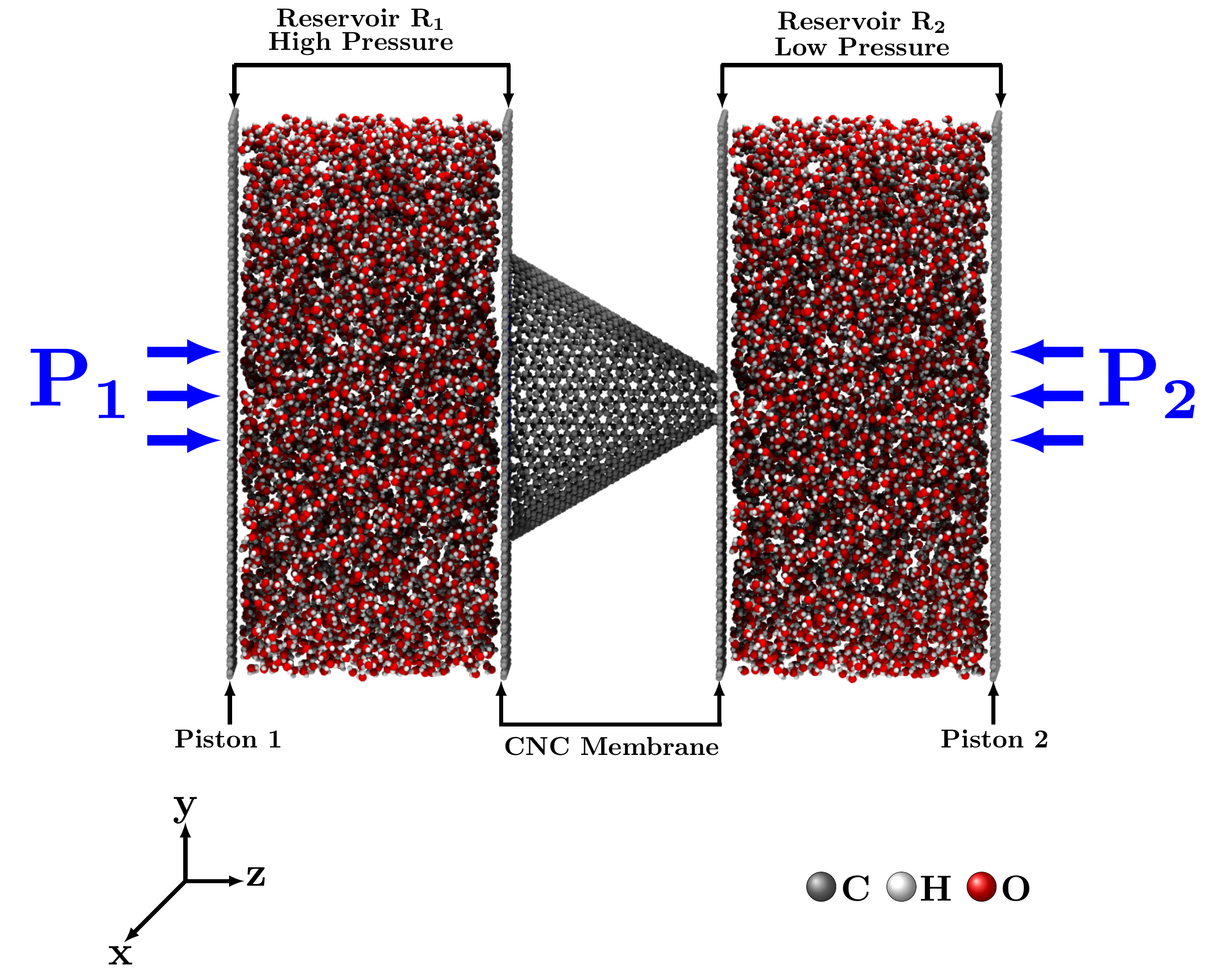}
	\end{center}
	\caption{The constructed system used in this work. A snapshot of the dual reservoir system shows the graphene pistons and the carbon nanocone membrane in the center. Graphene pistons 1 and 2 apply pressure P$_{1}$ and P$_{2}$ to the fluid. The blue arrows indicate the direction in which the pistons apply force to the water reservoirs, R$_{1}$ and R$_{2}$. The water flow was created by applying a pressure to reservoir R$_{1}$ that is greater than the pressure applied to reservoir R$_{2}$. The water molecules appear as a red oxygen atom connected to two white hydrogen atoms, and the carbon molecules in shades of gray.} 
	\label{fig_system}
\end{figure}

Each carbon nanocone is connected, through its large and small openings, to nanopores in two parallel sheets of graphene with lateral lengths of $L_{x}=L_{y}=10~nm$. The diameter of the nanopore  of each graphene sheet matches the diameter of one of the nanocone openings, so that the resulting systems consists of a conical hole in a bilayer graphene wall. The carbon nanocone axis was oriented parallel to the $z$ axis, while the nanopore sheets were perpendicular to the $z$ axis. Periodic boundary conditions were used in all directions. In doing so, the simulation box must be large enough in the $z$ direction to ensure that the molecules do not interact with each other across this boundary, so that the confined system is simulated correctly. We added two sheets of graphene with a length of $L_{x}=L_{y}=10~nm$ positioned at $10~nm$ on each side of the membrane opening that play the role of pistons, 1 and 2, applying pressure to the system. The pistons were oriented in the $xy$ plane, perpendicular to the $z$ axis. The $z$ positions of the carbon atoms in the pistons were allowed to move along the $z$ axis, while the relative position in the $xy$ plane was fixed. The water molecules were confined in the space between the pistons and the membrane to create the two water reservoirs R$_{1}$ and R$_{2}$. Each reservoir contains $5000$ molecules of water. Piston 1 applies pressure in the water reservoir R$_{1}$ against the CNC membrane. The forces applied to the pistons are evenly distributed equal to the pressures specified for each reservoir, allowing the location $z$ of the pistons to independently self-adjust to maintain the desired pressures and densities in each reservoir. During the simulation, it is important that the pistons remain well away from the inlet and outlet of the CNC membrane so as not to influence the flow through the channel. Pressures P$_{1}$ applied to piston 1 range from $200$ to $1000~bar$ in intervals of $200~bar$. The pressure P$_{2}$ applied to the piston 2 was $1~bar$. This pressure difference creates a favorable environment for water molecules to be forced to pass from the higher pressure reservoir R$_{1}$ to the lower pressure reservoir R$_{2}$. The difference in applied pressure generates a flow of water through the carbon nanocone. An external force applied to each carbon atom of the piston generated motion in the $z$ direction. The pressure exerted by the piston on the water was calculated using the equation $P=\frac{Fn}{A}$, where $P$ is the desired pressure applied by the piston, $F$ is the force applied to each atom in the direction $z$, $n$ is the number of carbon atoms in each piston, and $A$ is the surface area of the piston. The described system is shown in Figure~\ref{fig_system}.

The SHAKE~\cite{@10.1016/0021-9991(77)90098-5} algorithm is employed to stabilize the molecule bonds and angles. The nanocones and the graphene sheets are modeled by the Lennard-Jones potential (LJ) considering fixed bond lengths and angles~\cite{@10.1088/0959-5309} with effective carbon-carbon interaction energy $\epsilon_{CC}=0.086$~kcal$\cdot$mol$^{-1}$ and an effective diameter of $\sigma_{CC}=3.4$~\AA~\cite{@nature/35102535}. The carbon-oxygen energy $\epsilon_{CO}=0.11831$~kcal$\cdot$mol$^{-1}$ and the effective carbon-oxygen diameter $\sigma_{CO}=3.28218$~\AA~\cite{@nature/35102535}. The Lorentz-Berteloth mixing rules provided the LJ crossing parameters.

The parameters considered for the water model were defined and are shown in Table~\ref{tab_watermodels}. In this water model, the Lennard-Jones site is located on the oxygen atom, with parameters $\sigma$ and $\epsilon$. The charges of oxygen and hydrogen are $q_{O}$ and $q_{H}$, respectively. The TIP4P/2005 model places a negative charge $q_{M}$ at a point M at a distance $d_{OM}$ from the oxygen along the H-O-H bisector. The distance between the oxygen and hydrogen sites is $r_{OH}$. The angle formed between hydrogen, oxygen, and another hydrogen atom is given by $\theta_{HOH}$.

%%%%%%%%%%%%%%%%%%%%%%%%%%%%%%%%%%%%%%%%%%%%%%%%%%%%%%%%%%%
\begin{table}[H]
	\begin{center}
	\caption{Parameters of the TIP4P/2005 force field used for the water models. }
		\begin{tabular}{ c c }
			\hline
                $\epsilon_{OO}$ (kcal mol$^{-1}$) & 0.1852\\
                $\epsilon_{HH}$ (kcal mol$^{-1}$) & 0.0 \\
                $\sigma_{OO}$ (\AA) & 3.1589   \\
                $\sigma_{HH}$ (\AA) & 0.0   \\	
                $q_{O}$ (e) & 0.0   \\	
                $q_{H}$ (e) & 0.5564   \\	
                $q_{M}$ (e) & -1.1128  \\
                $d_{OM}$ (\AA) & 0.1546 \\	
                $r_{OH}$ (\AA) & 0.9572 \\		
                $\theta_{HOH}$ ($^{\circ}$) & 104.52 \\	
                \hline 
		\end{tabular}
		\label{tab_watermodels}
	\end{center}
\end{table}
%%%%%%%%%%%%%%%%%%%%%%%%%%%%%%%%%%%%%%%%%%%%%%%%%%%%%%%%%%%%%%%

The simulations were performed with the Large-scale Atomic/Molecular Massively Parallel Simulator (LAMMPS)~\cite{10.1006@jcph.1995.1039} package. We employ the Particle-Particle Particle-Mesh (PPPM) method to calculate long-range Coulomb interactions~\cite{10.1021@acs.jpcc.7b08326}. This method deals with the long-range interactions and the Coulomb field of real charges in a way that can interfere with its own images. We got around this problem by creating a simulation box on the $z$ axis around $500~nm$ for all systems, avoiding interaction with their own images and making it impossible to superimpose virtual images with real images, minimizing possible errors in the application of the method.

The simulation protocol involves the following steps:

\begin{enumerate}
    \item Pre-equilibrium in the NVE ensemble with a 0.5 ns MD run to minimize system energy keeping the pistons frozen (net force equal to zero). 
    \item Forces are applied in the pistons in order to impose 1 bar in each system to reach the water equilibrium densities at 300 K. Equilibration in the NPT ensemble during 1.0 ns. 
    \item  The pistons are frozen in the new equilibrium position. Equilibration in the NVT ensemble at 300 K controlled via the Nosé-Hoover thermostat~\cite{10.1080@00268978400101201} during 2.0 ns. 
    \item Nanopores are opened. Different forces are applied in each piston to mimic the pressure gradient. NPT ensemble during 10 ns at 300 K and different feed pressures.
\end{enumerate}

The system's dynamics features were evaluated by considering the flow rate calculations as given by the equation $\phi_{H_{2}O}=Av$, where $A$ is the graphene layer area ($3.4$x$3.4$~nm$^{2}$), and $v$ is the velocity of water flow acquired from the least-square linear regression line fitted to the data cloud which relates the mean molecular displacement along the tube axis as a function of time taken from the MD trajectory file.

%%%%%%%%%%%%%%%%%%%%%%%%%%%%%%%%%%%%%%%%%%%%%%%%%%%%%%%%%%%%%%%%%%%%%
\section{Results and discussion}

\begin{figure}[H]
	\begin{center}
		\includegraphics[width=6.in]{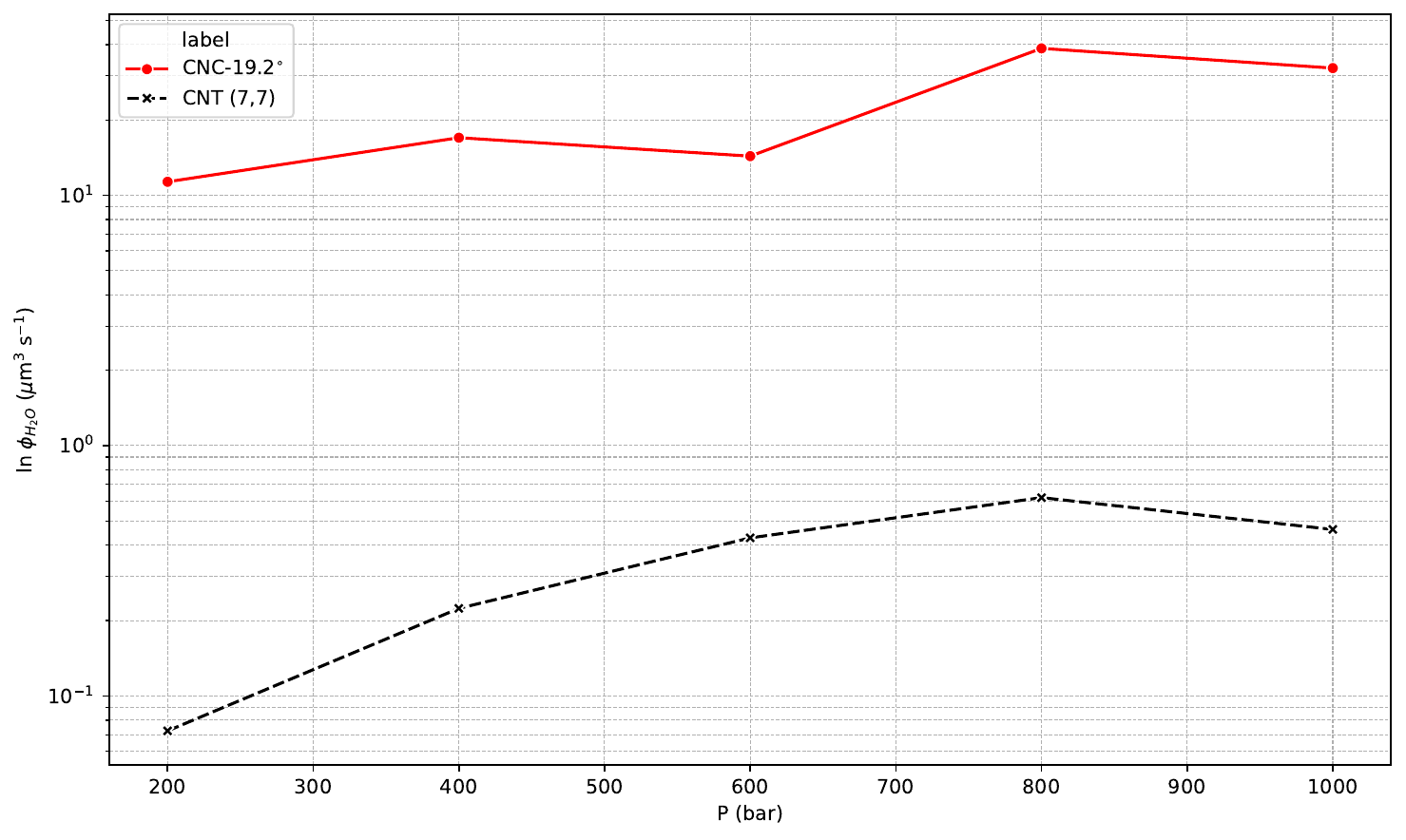}
	\end{center}
	\caption{Log of the flow rate as a function of the pressure gradient applied to the carbon nanocone CNC-19.2$^{\circ}$ and carbon nanotube ($7,7$).} 
	\label{fig-flux-log}
\end{figure}

Molecular dynamics simulations show that water flows through carbon nanocones significantly faster than through nanotubes of the same length and similar diameters, as we show in Fig. \ref{fig-flux-log}. We observed that the flow of water confined in carbon nanotubes is more uniform than in nanocones, and the water tends to organize itself into rows or chains, especially for nanotubes of small diameter. In nanocones, we observed flow magnitudes that are one order of magnitude larger than in nanotubes with similar structural properties \cite{@10.1021/acs.jpcb.3c02889}. The nanocone flow enhancement is driven by the conical shape itself, which creates a pressure gradient that naturally drives the water flow. The conical shape of the nanocones favors spontaneous pressure gradients and nearly frictionless sliding, promoting superflow of water - at a speed much higher than expected by classical hydrodynamic models.

We analyzed all experimentally characterized carbon nanocone structures \cite{@10.1016/0009-2614(94)00167-7,10.1038@41284}. Depending on the number of pentagons at the apex of the CNC~\cite{10.1016@S0038-1098(98)00210-5}, there are five symmetric CNCs with apex angles of CNC-112.9$^{\circ}$, CNC-84.6$^{\circ}$, CNC-60$^{\circ}$, CNC-38.9$^{\circ}$, and CNC-19.2$^{\circ}$. Here, the tips of the CNCs were cut (Fig. \ref{fig_cncs}) and their accessible pore areas are shown in Fig. \ref{fig_cncs} and described in Table \ref{tbl:cncs}.

First, we studied the water flow under different hydrostatic pressures in all CNCs. An approximately linear relationship between pressure and water flow was observed for all systems. Therefore, to obtain sufficient statistics in a limited simulation time, hydrostatic pressures from $200$ bar to $1000$ bar were adopted in the following simulations. Using these conditions, we examined the water flow performance of the systems shown in Fig. \ref{fig_system}.

\begin{figure}[H]
	\begin{center}
		\includegraphics[width=6.in]{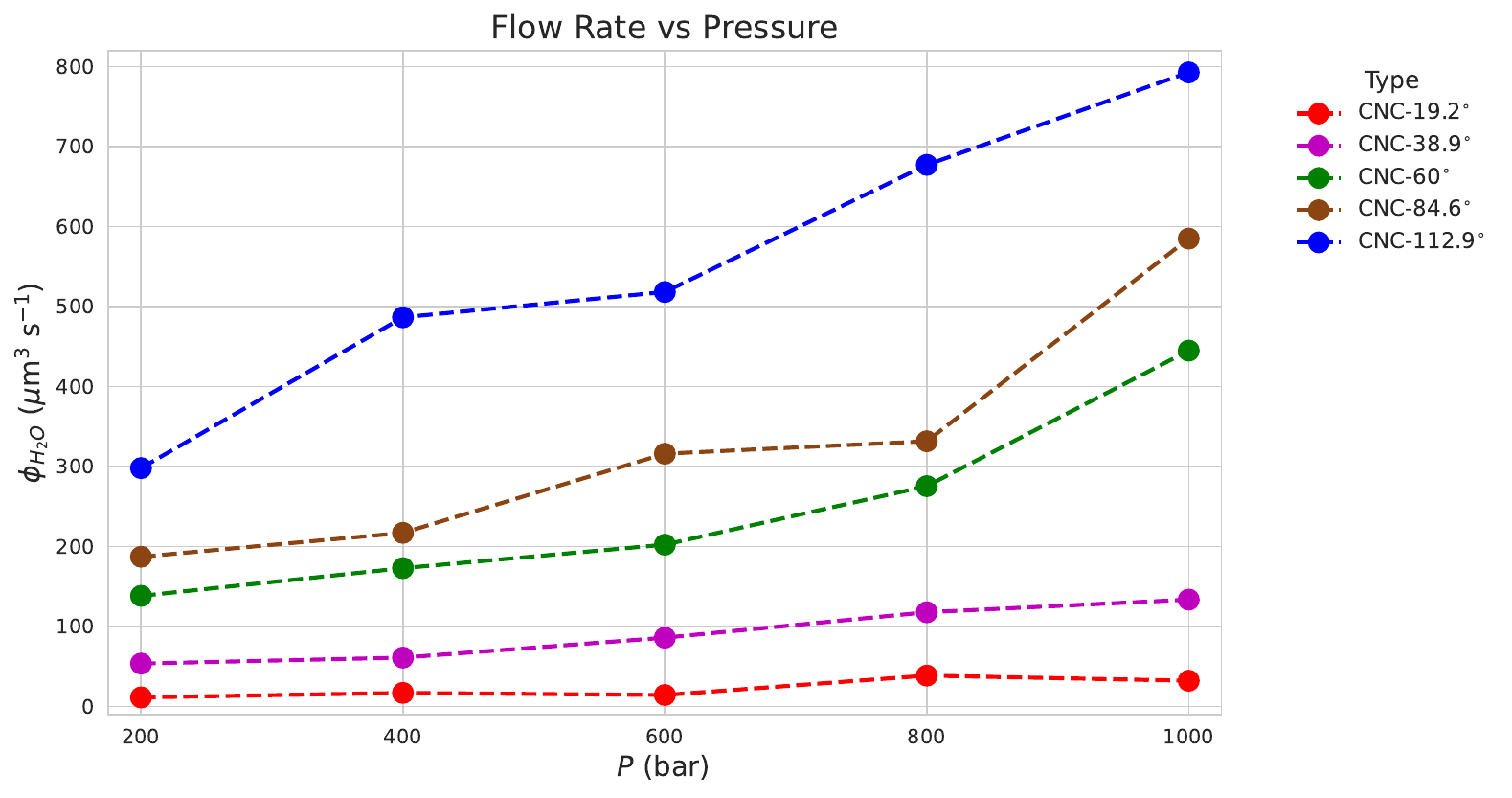}
	\end{center}
	\caption{Flow rate as a function of the pressure gradient applied to the carbon nanocones.} 
	\label{fig-flux}
\end{figure}

The calculated flow rates are shown in Fig. \ref{fig-flux} as a function of pressure gradient, for several cone structures. The floW rate increases enormously with cone apex angle, resulting in increases of more than one order of magnitude for a given pressure gradient. When water flows in CNCs from the larger opening to the smaller opening, the mobility of water molecules increases smoothly, allowing more water to enter, which contributes to their higher water flux as the CNC opening increases. Similarly, when water flows in the opposite direction, it can easily diffuse out, which also provides relatively lower flows. The larger pores on the base side in CNCs and the corresponding higher probability of collecting water from the solution under pressure are possibly responsible for the higher water fluxes going from the larger diameter to the smaller one, which explains the obtained results.

\begin{figure}[H]
	\begin{center}
		\includegraphics[width=6.4in]{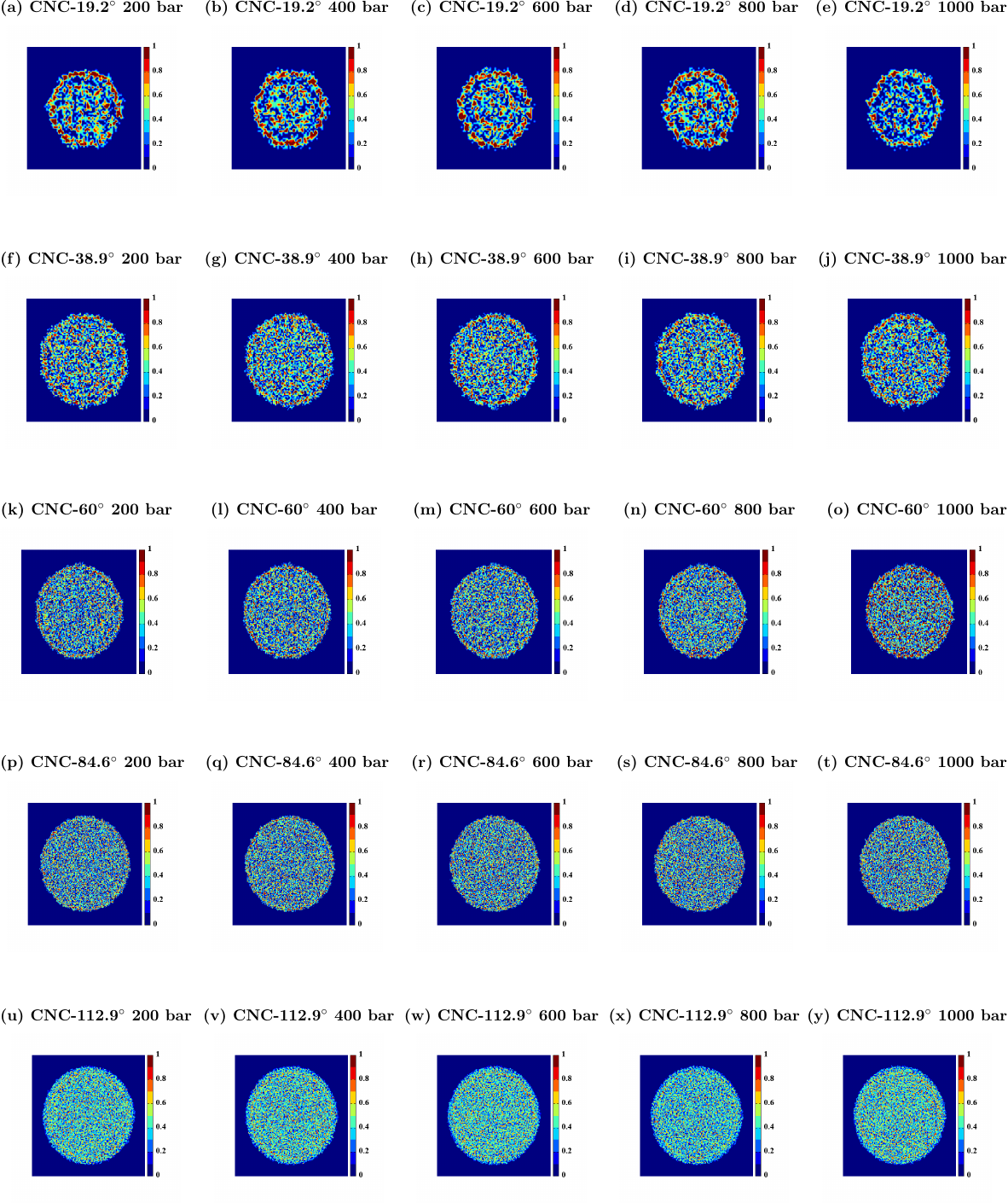}
	\end{center}
	\caption{Density maps in the $xy$ direction for the carbon nanocones. Dark blue regions have a low probability of finding water molecules, while red regions have a high probability of finding water molecules.} 
	\label{fig-maps-1}
\end{figure}

\begin{figure}[H]
	\begin{center}
		\includegraphics[width=6.4in]{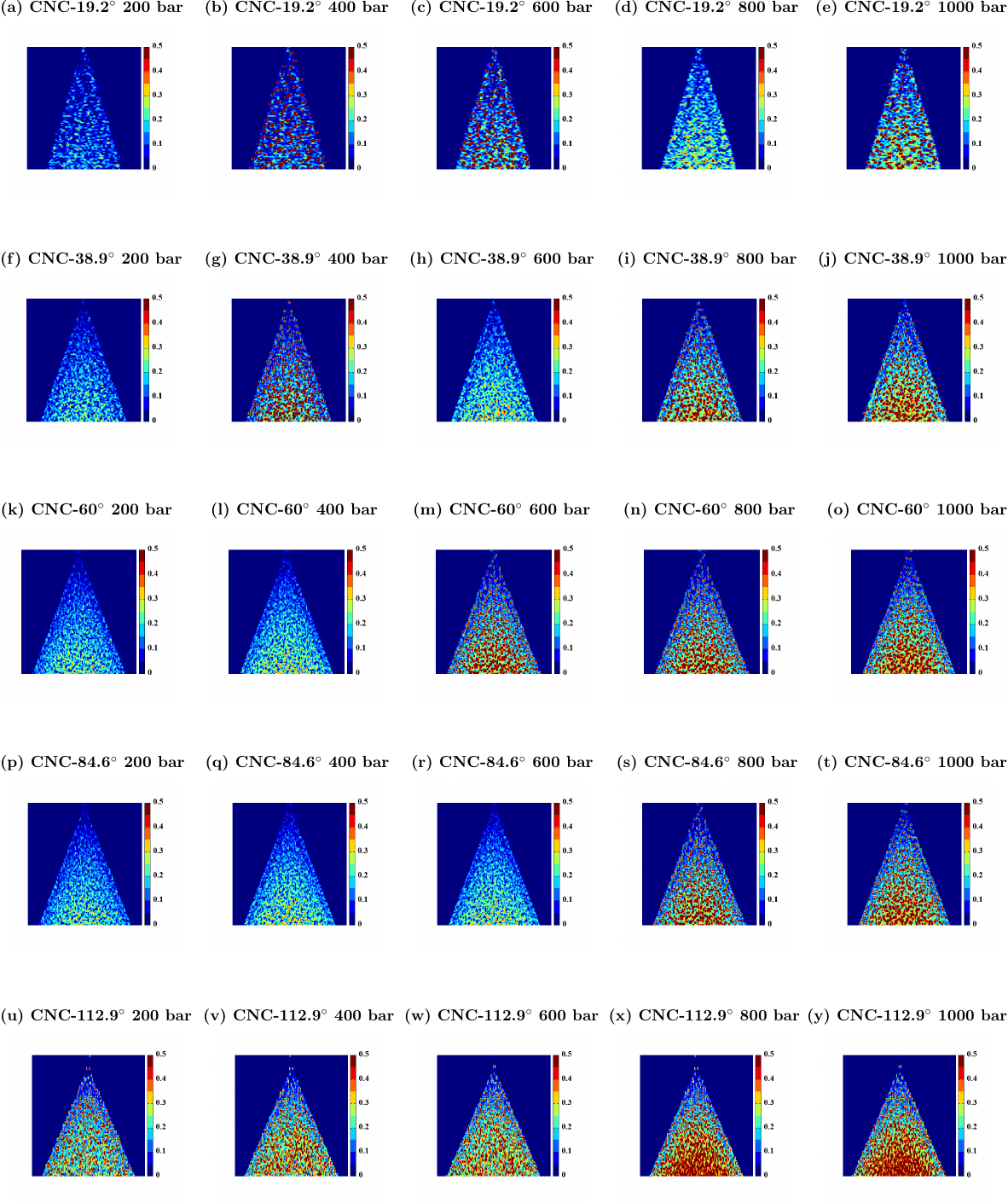}
	\end{center}
	\caption{Density maps in the $yz$ direction for the carbon nanocones. Dark blue regions have a low probability of finding water molecules, while red regions have a high probability of finding water molecules.} 
	\label{fig-maps-2}
\end{figure}

Figs. \ref{fig-maps-1} and \ref{fig-maps-2} show the water density maps of the simulated systems. In the CNC-19.2$^{\circ}$ system, the water is relatively ordered (Fig. \ref{fig-maps-1} (a-e) and Fig. \ref{fig-maps-2} (a-e)), and consequently the water fluxes in this CNC are small. When the apex angle is increased, the water molecules become less ordered, as in bulk water, causing an increase in their flux. We can observe this from CNC-38.9$^{\circ}$ onwards.\cite{10.1016@j.carbon.2017.12.039} 

It should be noted that in our simulations, all carbon atoms in CNC channels are electroneutral. However, in a real system, carbon atoms at the edge of CNC channels are prone to be oxidized and develop small net charges. The net charges can polarize the passing water molecules and cause the water molecules to reorient. Therefore, we beleive that the small net charges and the corresponding ordered water structure can further enhance the water flux~\cite{10.1021@acs.jpcc.7b04283,10.1021@la4018695}

\begin{figure}[H]
	\begin{center}
		\includegraphics[width=6.4in]{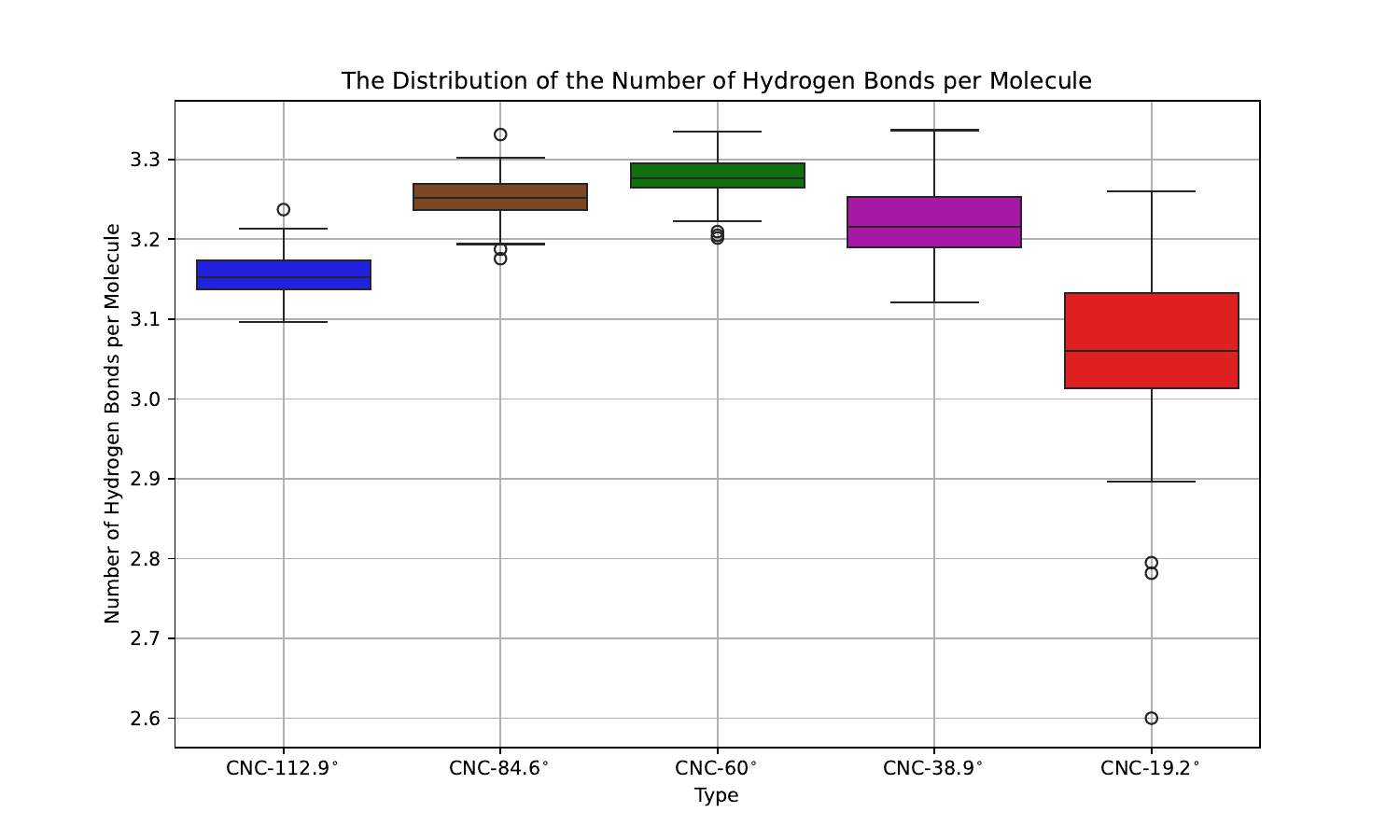}
	\end{center}
	\caption{The distribution of the number of hydrogen bonds per molecule as boxplot and by each carbon nanocones structure.} 
	\label{fig-1-distribution-hb}
\end{figure}

We also investigated the dependence of water transport on one main aspect: the hydrogen bonding behavior of confined water molecules. Fig. \ref{fig-1-distribution-hb} shows that the absolute number of hydrogen bonds per molecule within the nanocone does not vary substantially across different structures. However, a more detailed assessment revealed that each nanocone geometry exhibits a distinct number of broken hydrogen bonds relative to the entry region.

\begin{figure}[H]
	\begin{center}
		\includegraphics[width=6.4in]{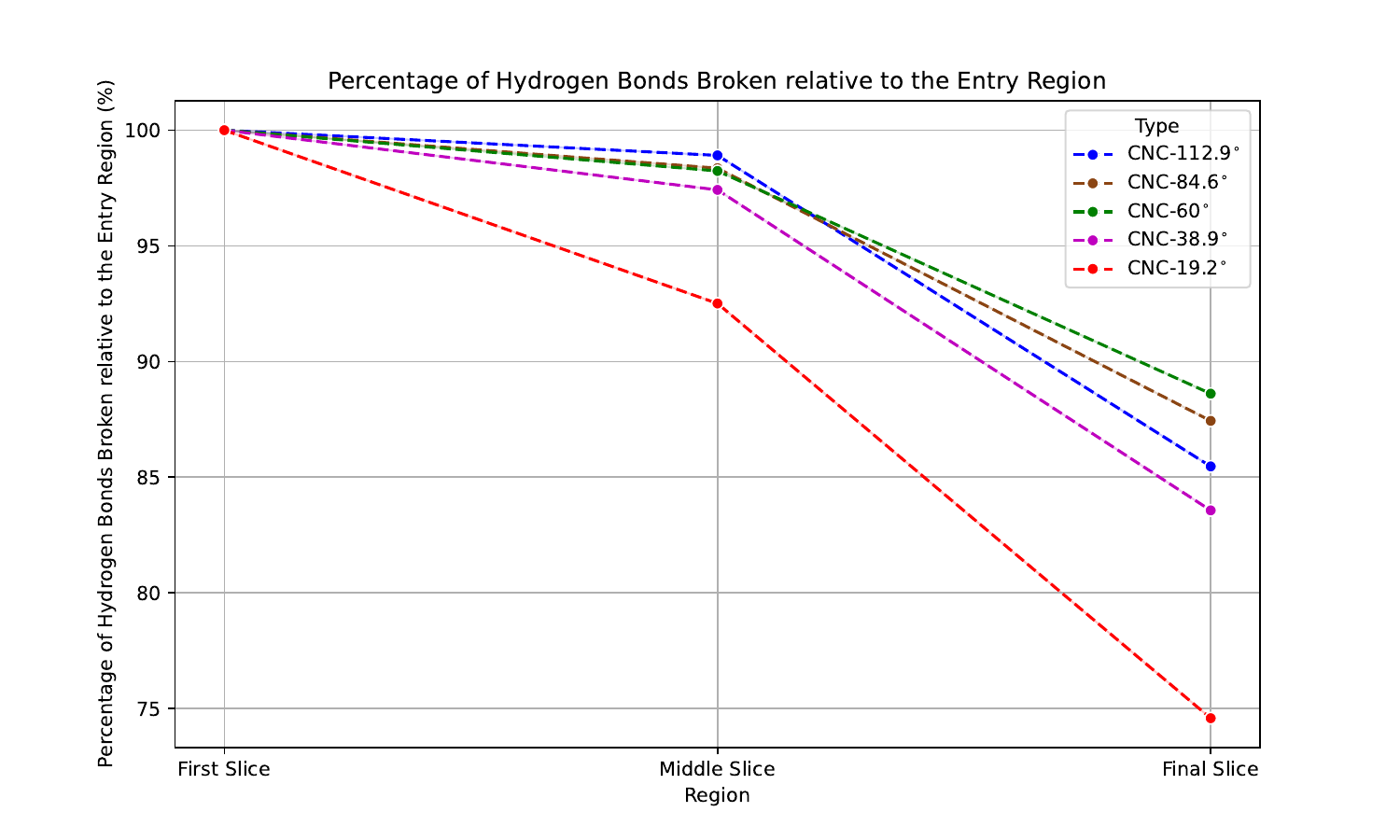}
	\end{center}
	\caption{The percentage of hydrogen bonds broken relative to the carbon nanocones entry region for each carbon nanocones structure.} 
	\label{fig-4-hb-broken}
\end{figure}

The percentage of hydrogen bonds broken were obtained by dividing each nanocone into three equally high sections—namely, the entry (first slice), the middle (mille slice), and the exit (final slice) regions. As shown in Fig. \ref{fig-4-hb-broken}, when comparing the number of hydrogen bonds in the middle and final slices to those in the entry region, structure CNC-19.2$^{\circ}$ presents a substantially larger drop in hydrogen bonds. This steep decline indicates that water molecules traversing CNC-19.2$^{\circ}$ must, on average, break a greater number of hydrogen bonds to move from one region to the next. Such enhanced bond disruption likely translates into a higher energy barrier for molecular transport, thereby contributing to the lower flux in the CNC-19.2$^{\circ}$ configuration.

In summary, our simulations reveal that while the overall hydrogen bonding network of confined water does not dramatically differ among the CNC structures, the interplay within the local hydrogen bond dynamics is crucial. The higher number of hydrogen bonds broken along the structure CNC-19.2$^{\circ}$ result in a more rigid and structured water network that, in turn, impedes flux. Conversely, structures with lower percentage of hydrogen bonds broken presents more gradual change in hydrogen bonding along the cone and that allow water molecules to traverse the nanocone with less disruption. These insights provide a intereseting understanding of how nanoscale geometric parameters can be manipulated to optimize fluid transport in confined systems, which is pivotal for the design of efficient nanofluidic devices.

%%%%%%%%%%%%%%%%%%%%%%%%%%%%%%%%%%%%%%%%%%%%%%%%%%%%%%%%%%%%%%%%%%%%%
\section{Conclusions}
In this work, molecular dynamics simulations were conducted to investigate the performance of confined water flow in CNC membranes. The results show that CNCs can closely optimize water flow depending on the diameter of the CNC. In addition, CNC with an apex angle of 19.2$^{\circ}$ provides the lowest water flow, due to a greater ordering of the water present in the channel, a fact that is linked to the ability of water to form a larger hydrogen bond network in typical systems with diameters of this size, obtaining a single-layer water structure. Under pressure, CNCs also exhibit better water flow performance as we increase the apex angle of the CNCs, for these systems we have a typical water mobility behavior of bulk water. An application of these systems would be the fabrication of membranes to be used in water desalination processes. Water flow in CNCs has enormous technological potential because of its unique properties of superflux, extreme confinement, and low friction. Therefore, the fabrication of CNC membranes with uniform configuration and precise pore sizes for desalination will become a reality in the near future. Carbon nanocones are becoming a promising platform for clean, smart, and efficient technologies. The combination of transport speed, selectivity, and structural control put them ahead of other nanostructures for various purposes. This work offers great promise for the next generation of desalination devices.

%%%%%%%%%%%%%%%%%%%%%%%%%%%%%%%%%%%%%%%%%%%%%%%%%%%%%%%%%%%%%%%%%%%%%
%% The "Acknowledgement" section can be given in all manuscript
%% classes.  This should be given within the "acknowledgement"
%% environment, which will make the correct section or running title.
%%%%%%%%%%%%%%%%%%%%%%%%%%%%%%%%%%%%%%%%%%%%%%%%%%%%%%%%%%%%%%%%%%%%%
\begin{acknowledgement}

This work is funded by the Brazilian scientific agency Conselho Nacional de Desenvolvimento Científico e Tecnológico (CNPq), Fundação de Amparo à Pesquisa do Estado da Bahia (FAPESB), and the Brazilian Institute of Science and Technology (INCT) in Carbon Nanomaterials with collaboration and computational support from Universidade Federal de Minas Gerais (UFMG) and Universidade Federal da Bahia (UFBA). BHSM thanks Rede Mineira de Materiais Bidimensionais (REDE 2D) and the science agency Fundação de Amparo à Pesquisa do Estado de Minas Gerais (FAPEMIG) for financial support. EEM appreciates Edital PRPPG 010/2024 Programa de Apoio a Jovens Professores(as)/Pesquisadores(as) Doutores(as) - JOVEMPESQ Project 24460. Finally, authors acknowledge the National Laboratory for Scientific Computing (LNCC/MCTI, Brazil) for providing HPC resources of the SDumont supercomputer, which have contributed to the research results reported within this paper. URL: \url{http://sdumont.lncc.br}.

\end{acknowledgement}

%%%%%%%%%%%%%%%%%%%%%%%%%%%%%%%%%%%%%%%%%%%%%%%%%%%%%%%%%%%%%%%%%%%%%
%% The same is true for Supporting Information, which should use the
%% suppinfo environment.
%%%%%%%%%%%%%%%%%%%%%%%%%%%%%%%%%%%%%%%%%%%%%%%%%%%%%%%%%%%%%%%%%%%%%
%\begin{suppinfo}

%This will usually read something like: ``Experimental procedures and
%characterization data for all new compounds. The class will
%automatically add a sentence pointing to the information on-line:

%\end{suppinfo}

%%%%%%%%%%%%%%%%%%%%%%%%%%%%%%%%%%%%%%%%%%%%%%%%%%%%%%%%%%%%%%%%%%%%%
%% The appropriate \bibliography command should be placed here.
%% Notice that the class file automatically sets \bibliographystyle
%% and also names the section correctly.
%%%%%%%%%%%%%%%%%%%%%%%%%%%%%%%%%%%%%%%%%%%%%%%%%%%%%%%%%%%%%%%%%%%%%
\bibliography{achemso-demo}

\end{document}